\documentclass[journal,10pt,twocolumn]{IEEEtran}
\usepackage{amsmath,epsfig,stfloats,amssymb,subfigure}
\usepackage{cite}
\usepackage{psfrag,color}
\usepackage{multirow}
\usepackage{enumerate,soul}
\def\x{{\mathbf x}}

\usepackage{setspace}

\begin{document}

\title{Device-Free Person Detection and Ranging \\in UWB Networks}

\author{Yakup~Kilic,~\IEEEmembership{Student~Member,~IEEE,}
        Henk~Wymeersch,~\IEEEmembership{Member,~IEEE,}
        Arjan~Meijerink,~\IEEEmembership{Senior~Member,~IEEE,}
        Mark~J.~Bentum,~\IEEEmembership{Senior~Member,~IEEE,}
        and~William~G.~Scanlon,~\IEEEmembership{Member,~IEEE}
\thanks{Y. Kilic, A.~Meijerink, M.~J.~Bentum, and W.~G.~Scanlon are are with the Telecommunication Engineering Group, University of Twente, 7500~AE Enschede, The Netherlands (e-mail: \{y.kilic, a.meijerink, m.j.bentum, w.g.scanlon\}@utwente.nl). H. Wymeersch is with the Department of Signals and Systems, Chalmers University of Technology, Gothenburg, Sweden (e-mail:henkw@chalmers.se). W.~G.~Scanlon is also with the School of Electronics, Electrical Engineering and Computer Science, The Queen’s University of Belfast, Belfast BT3 9DT, U.K.
(e-mail:w.scanlon@qub.ac.uk).
This research was supported, in part, by the European Research Council, under Grant No. 258418 (COOPNET). }}

\maketitle

\begin{abstract} We present a novel device-free stationary person detection and ranging method, that is applicable to ultra-wide bandwidth (UWB)  networks. The method utilizes a fixed UWB infrastructure and does not require a training database of template waveforms. Instead, the method capitalizes on the fact that a human presence induces small low-frequency variations that stand out against the background signal, which is mainly affected by wideband noise. We analyze the detection probability, and validate our findings with numerical simulations and experiments with off-the-shelf UWB transceivers in an indoor environment.
\end{abstract}
\section{Introduction}\label{sec:introduction}
\PARstart{W}{ireless} localization and tracking has attracted a great deal of research interest from the research community, as location-awareness is fast becoming an essential feature in many application areas \cite{SunChen05}. For indoor scenarios, ultra-wide bandwidth (UWB) transmission is a promising technology, due to its high-resolution ranging and obstacle penetration capabilities \cite{Joon-Yong_Lee,Falsi,Jourdan,Dardari,Gezici1}. Most practical UWB localization systems rely on targets (e.g., objects, people) to carry an active UWB device, which is used to facilitate time-difference-of-arrival or time-of-arrival measurements\cite{Gezici1}. In some scenarios (e.g., intruder detection, elderly care, smart environments, emergency response) it is desirable to have the ability to track people and assets in a passive manner, without requiring them to be equipped with any radio-frequency (RF) device. This is known as \emph{device-free} localization, an overview of which can be found in the survey papers \cite{Teixeira,Deak}. Traditional device-free localization techniques were vision-based, relying on infrared motion detectors and video camera surveillance, but are limited to visible line-of-sight (LOS).
Modern techniques overcome this problem through RF-based transmission, where received RF-signals  are affected by the presence of people or assets in a quantifiable way. Research in this area can be broadly differentiated based on the narrow-band and wide-band nature of the signals involved.

The earliest works in narrow-band device-free localization \cite{Woyach,Youssef,Zhang} exploited human body-induced shadowing on the transmission links by measuring received signal strength indicators (RSSI). In particular, \cite{Woyach} detects  motion and estimates velocities based on variations in RSSI, while \cite{Youssef,Zhang} employs filters or thresholds to detect changes in the RSSI. More recent works \cite{Xu,Kosba} provided improvements, by partitioning the environment into cells and employing discriminant analysis in each cell \cite{Xu} and non-parametric statistical anomaly detection \cite{Kosba}. In \cite{Patwari1,Patwari2,Wilson1,Zhao,Savazzi}, received signal strength (RSS) variance is considered in an effort to improve robustness against environmental changes. Aspects of mobility on RSS variance were considered in \cite{Zhao,Wilson1,Savazzi}, while models allowing distinction between mobile and static people were treated in \cite{Wilson2,Wilson3}, and for single and multiple people in  \cite{Nannuru}. The term \emph{radio tomographic imaging} was introduced in \cite{Wilson2}, as a general description  for these systems, due to their similarity with computed tomography. While the low-cost implementation of narrow-band radios is quite attractive, susceptibility of the system to multi-path fading makes it hard to develop accurate models for dense, cluttered environments. These drawbacks can be overcome by considering larger RF bandwidths.

UWB was demonstrated as an effective technique for human-being detection through respiratory movement in \cite{Yarovoy,Ossberger}. Recently, experimental demonstrations were also given for different positions of the body and different antenna polarizations\cite{Nezirovic}, MIMO UWB\cite{Yanyun_Xu}, and sensing of the person through obstructions\cite{Sachs,Li}. For the estimation of respiration and heart rates, an analytical framework and a frequency domain technique were developed for single person in \cite{Venkatesh}, experimental results were given for multiple targets in \cite{Rivera}, and related Cram\'{e}r--Rao lower bounds were calculated in \cite{Gezici2}. A new time-variant channel response model is introduced for breathing detection and human target ranging in \cite{Casadei}, and the harmonics and the intermodulation between respiration and heart signals are analyzed in \cite{Lazaro}. While these works dealt with the detection of static people from  breathing information, human-body detection and tracking were also studied experimentally for moving people in an open area in \cite{Chang1,Chang2}. Passive object detection and tracking was also studied in UWB  networks, in which an initial study on tracking was performed in \cite{Chang3}, which derived Cram\'{e}r-Rao lower bounds, assuming a specular reflection model. Imaging of environments and objects based on a single UWB transmission was considered in \cite{Thoma,Guo}, and extended to multiple receivers in \cite{Paolini,Chiani,Bartoletti,Muharrem}. This has resulted in a flurry of research papers, investigating the impact of system geometry  \cite{Paolini}, optimum detection metrics \cite{Chiani}, imaging methods \cite{Muharrem}, and signal analysis methods \cite{JSheng}. While UWB device-free localization is able to overcome many of the drawbacks of narrowband signals, a number of issues are still unresolved. First of all, although experimental demonstrations were given for detecting the presence of people via UWB transmission, there is no unified signal processing technique for device-free detection, ranging, and localization of people. Second, in most existing methods, there is a need to develop a training database prior to localization. This database needs to be updated frequently, as changes in the environment may occur, precluding fast deployment.

In this paper, we develop a novel device-free indoor human-body detection and ranging method, that can be applicable to UWB networks and used as a complementary method to existing device-free or device-based localization methods. Our detection method operates in the time domain unlike the works \cite{Venkatesh,Rivera,Casadei} and \cite{Pratt}, in which frequency domain and polarization-based techniques are considered, respectively. Our method is based on the observation, corroborated by measurements, that a stationary human body introduces small temporal variations in the received UWB signal, even when the person is standing still. As such, it does not require any training.\footnote{In practice, some calibration may be required to determine the ambient noise power and to set the system operating parameters appropriately.} Since the multi-path UWB signal, reflected from  other objects is stationary over relatively long time windows, the deployment of this system in dense, cluttered environments does not affect the performance. Based on this observation, we describe a corresponding UWB signal and develop a generic detection technique. This detection technique is validated by numerical simulations and experimental measurements. Finally, we provide a limited set of localization results, based on experimental data.

The remainder of this paper is organized as follows. In Section \ref{sec:Problem-formulation}, we formulate the problem.  In Section \ref{sec:Detection}, we describe a detection criterion, and analyze the corresponding false alarm and missed detection probabilities in Section \ref{sec:Performance}. Simulations and experimental results are provided in Section \ref{sec:Performance Evaluation}, before we draw conclusions in Section~\ref{sec:Conclusion}.

\section{Problem Formulation}
\label{sec:Problem-formulation}

\subsection{Localization System}
We consider a system with $N_{\mathrm{a}}$ UWB radios (called beacon or anchor nodes), with a priori known positions   $\mathbf{x}_i \in \mathbb{R}^2$ and a passive, human target with an unknown position $\mathbf{x} \in \mathbb{R}^2$. We assume that the anchors exchange signals which reflect off the target. Based on such a reflected signal from anchor $i$ to anchor $j$, we obtain an estimate of the time of flight, or a corresponding distance
\begin{eqnarray}\label{eq:range}
\hat{d}_{i,j} = ||\mathbf{x}-\mathbf{x}_{i}||+||\mathbf{x}-\mathbf{x}_{j}|| + e_{i,j},
\end{eqnarray}
where $e_{i,j}$ is measurement error. This enables the system to approximately locate the target on an ellipse  whose foci are $\mathbf{x}_i$ and $\mathbf{x}_j$, and with the length of the major axis equal to $\hat{d}_{i,j}$. Given a collection of such measurements, the least-squares estimate of $\mathbf{x}$ is
\begin{eqnarray}\label{eq:mlPositionEstimate}
\hat{\mathbf{x}}=\arg \min_{\mathbf{x}}\sum_{(i,j)} \left(\hat{d}_{i,j}-||\mathbf{x}-\mathbf{x}_{i}||-||\mathbf{x}-\mathbf{x}_{j}||\right)^{2},
\end{eqnarray}
where the summation goes over all pairs of transmitters $i$ and receivers $j$ that have a measurement. Our goal is to determine $\hat{d}_{i,j}$ based on human body-induced signal variations, and without a template waveform.

\subsection{Signal Model}

\begin{figure}[!t]
\begin{center}
\hskip -0.5cm
\psfrag{xlabel}{\footnotesize{delay [ns]}}
\psfrag{ylabel}{\hspace{-15mm}\footnotesize{amplitude [ADC counts]}}
\psfrag{withbodyxxxxxx}{\footnotesize{person present}}
\psfrag{withoutbody}{\footnotesize{person absent}}
\includegraphics[width=1.04\columnwidth]{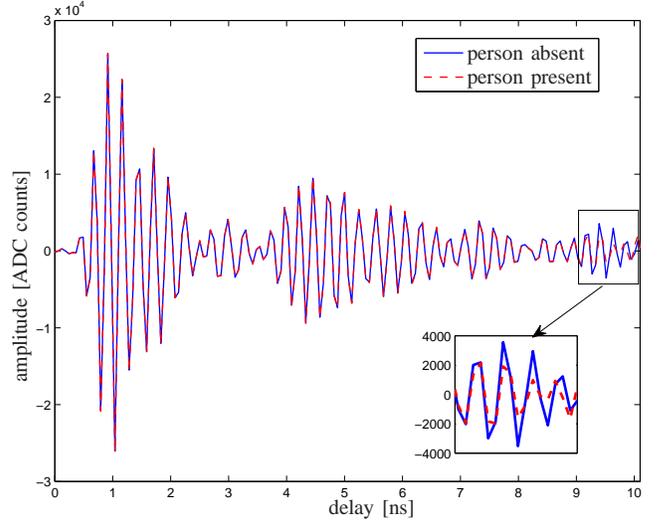}
\end{center}
\caption{Mean of $100$ UWB measurements, taken continuously within 20 seconds in an indoor LOS environment, in the presence and absence of a person. }
\label{fig:receivedWaveformFixedTime}
\end{figure}

\begin{figure}[!t]
\begin{center}
\psfrag{xlabel}{\footnotesize{time [s]}}
\psfrag{ylabel}{\hspace{-15mm}\footnotesize{amplitude [ADC counts]}}
\psfrag{withbodyxxxxxxx}{\footnotesize{person present}}
\psfrag{withoutbody}{\footnotesize{person absent}}
\includegraphics[width=1\columnwidth]{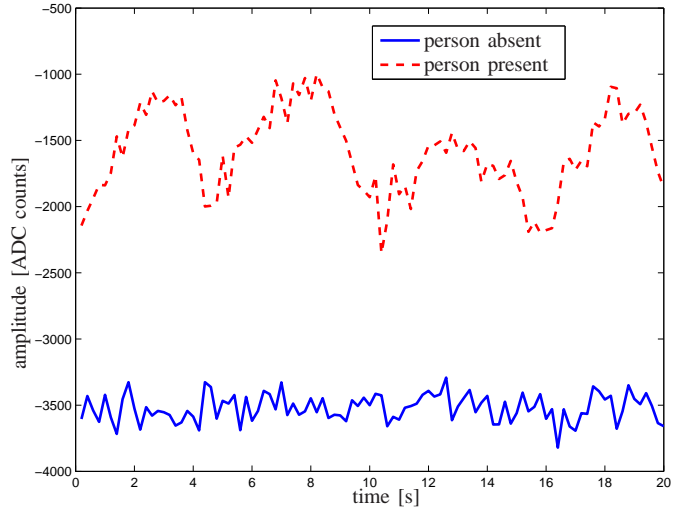}
\end{center}
\caption{Variation of the received waveform over time at delay instant 9.5~ns, in the presence and absence of a person.}
\label{fig:receivedWaveformFixedTauEmpty}
\end{figure}

In this section, we will develop a signal model that captures the effect of a human presence on UWB signals. Let us consider the following scenario\footnote{A detailed description of the experimental setup is given in Section~\ref{sec:Experimental-Results}.}: a transmitter and a receiver, separated by a distance of five meters in an indoor LOS environment, and a person on an ellipse around transmitter and receiver with major axis length 7.7 m (or, equivalently, a delay of 9~ns with respect to the direct signal path). Fig.~\ref{fig:receivedWaveformFixedTime} shows  received UWB waveforms when the person is present (dashed line) or absent (solid line). We observe that the person affects the signal to some small extent around 9 ns after the first arrival. However, without a template (in this case the solid line) it is unclear how to determine the delay related to the person. A different view is offered in Fig.~\ref{fig:receivedWaveformFixedTauEmpty}, showing the evolution of the signal at a fixed delay of 9.7 ns, in the presence and absence of a person. Such a signal is obtained by receiving a periodic waveform, and aligning the copies in the delay domain. As one would expect, in the absence of a person, the signal is essentially constant (here around -3500 analog-to-digital converter (ADC) counts) and affected by noise. Interestingly, in the presence of a person, the signal evolves slowly over time, due to minor temporal variations, e.g., induced by breathing.

Based on these observations, we pose the following model. We transmit $N_{\mathrm{rep}}$ copies, with a period $T_{\mathrm{rep}}$, of a ranging signal\cite{Gezici1},
\begin{align}
s(t)&= \sqrt{\frac{E_{\mathrm{s}}}{N_{\mathrm{f}}}}\sum_{j=0}^{N_{\mathrm{f}}-1}b_{j}p(t-jT_\mathrm{f}-c_{j}T_{\mathrm{c}}),  \label{eq:transmittedSignalModel}
\end{align}
where $E_{\mathrm{s}}$ denotes the energy of the signal, $N_{\mathrm{f}}$ is the number of pulses in $s(t)$, $b_{j}\in\{\pm1\}$ denotes the polarity code, $T_{\mathrm{f}}$ is the duration of the frame and $p(t)$ denotes the unit energy UWB pulse of duration $T_{\mathrm{p}}<T_{\mathrm{f}}$, $c_{j}\in\{0, 1,\ldots, N_{\mathrm{h}}-1\}$ is the time-hopping code, $T_{c}$ and $N_{\mathrm{h}}$ are the chip duration and the number of chips per frame, respectively. The total duration of the signal is $T_{\mathrm{s}}=N_{\mathrm{f}}T_{\mathrm{f}}$, being smaller than the period $T_{\mathrm{rep}}$.
The receiver coherently combines $N_{\mathrm{f}}$ pulses during each of $N_{\mathrm{rep}}$ repetitions, leading to the following received signal
\begin{align}
r_{\mathrm{rep}}(t)&= \sqrt{E_{\mathrm{s}}}\sum_{k=1}^{N_\mathrm{rep}}\sum_{l=1}^{L}\alpha_{l}p(t-\tau_{l}-k T_{\mathrm{rep}})\nonumber \\ & + \sqrt{E_{\mathrm{s}}} \sum_{k=1}^{N_\mathrm{rep}}\sum_{l=1}^{M}\alpha_{\mathrm{p},l}(t)p(t-\tau_{\mathrm{p},l}-k T_{\mathrm{rep}})+n(t), \label{eq:receivedSignalModel}
\end{align}
where the first term comprises $L$ signal paths, being irrespective of the presence of the person, in which $\alpha_{l}$ and $\tau_{l}$ are the channel coefficient
and the delay of the $l$-th discrete path, respectively. Assuming the delays are ordered, $\tau_{1}$ corresponds to the direct-LOS path. The second term in (\ref{eq:receivedSignalModel}) corresponds to the person, where in view of our model, there are $M$ channel coefficients $\alpha_{\mathrm{p},l}(t)$ varying slowly over time due to the effects of the human body. Among them, $\alpha_{\mathrm{p},1}(t)$ represents the signal path that is directly reflected off the person (i.e., direct reflection), whereas the remaining ones are related to signal paths that are reflected off the person and the other reflection sources in the environment (i.e, indirect reflections). Note that the delay associated with the direct reflected path $\tau_{\mathrm{p},1}$ is always smaller than the delays associated with the indirect reflected paths $\tau_{\mathrm{p},l}$ for $l=2,\ldots,M$, since the direct reflected path always travels a shorter distance. The noise $n(t)$ is assumed to be zero-mean independent and identically distributed Gaussian noise with power spectral density $N_{0}/2$.

Direct sampling of $r_{\mathrm{rep}}(t)$ at a sufficiently high rate $W$ over a window $T_{\mathrm{delay}}$,
and aligning the $N_{\mathrm{rep}}$ copies, we obtain a two-dimensional model with $r(kT_{\mathrm{rep}},m/W)=r_{\mathrm{rep}}(kT_{\mathrm{rep}}+m/W)$, $\alpha_{\mathrm{p},l}(kT_{\mathrm{rep}},m/W)$ $=$ $\alpha_{\mathrm{p},l}(kT_{\mathrm{rep}}+m/W)$ and $n(kT_{\mathrm{rep}},m/W)=n(kT_{\mathrm{rep}}+m/W)$ for $k=1,2,\ldots,N_{\mathrm{rep}}$ and $m=0,1,\ldots,T_{\mathrm{delay}}W-1$:
\begin{align}
r(kT_{\mathrm{rep}},m/W)&=\sqrt{E_{\mathrm{s}}}\sum_{l=1}^{L}\alpha_{l}p(m/W-\tau_{l})\nonumber \\ & + \sqrt{E_{\mathrm{s}}}\sum_{l=1}^{M}\alpha_{\mathrm{p},l}(kT_{\mathrm{rep}},m/W)p(m/W-\tau_{\mathrm{p},l}) \nonumber \\
& +n(kT_{\mathrm{rep}},m/W),\label{eq:2DreceivedSignalModel}
\end{align}
with
\begin{align} \label{eg:noiseEquation}
\mathbb{E}\left\{ n(kT_{\mathrm{rep}},m/W)n(k'T_{\mathrm{rep}},m'/W)\right\}  = \nonumber \\ \frac{N_{0}W}{2}\delta(k-k')\delta(m-m'),
\end{align}
where $\mathbb{E}\{.\}$ denotes the statistical expectation and $\delta(.)$ is the discrete delta function. In view of our model,
the second term in (\ref{eq:2DreceivedSignalModel}) again corresponds to the person, which introduces $M$ channel coefficients that vary slowly over time (i.e., as a function of $k$).  We can further break up the signal corresponding to the person in a static part (i.e, independent of $k$) and a slowly-varying part as
\begin{align}\label{eq:fixedTimeVarying}
&\sqrt{E_{\mathrm{s}}}\alpha_{\mathrm{p},l}(kT_{\mathrm{rep}},m/W)p(m/W-\tau_{\mathrm{p},l}) = \nonumber \\ &\sqrt{E_{\mathrm{s}}}C(m/W-\tau_{\mathrm{p},l})+f(\tau_{\mathrm{p},l})w(kT_{\mathrm{rep}})V(m/W-\tau_{\mathrm{p},l})
\end{align}
where for a fixed $m$, $C(m/W-\tau_{\mathrm{p},l})$ is a constant, depending on the propagation effects of the human body and the received signal power, $f(\tau_{\mathrm{p},l})$ is a decreasing function in $\tau_{\mathrm{p},l}$ to model the effect of propagation loss on the time-varying signal part, and
\begin{eqnarray}
f^2(\tau_{\mathrm{p},l}) =  E_sG(\tau_{\mathrm{p},l}),\label{eq:f_tau_zero_square}
\end{eqnarray}
where $G(\tau_{\mathrm{p},l})$ refers to the path gain\footnote{The distance dependence of the received signal power is typically expressed by path-gain models. Given a reference received signal power $P_{\mathrm{r},0}$ at a reference distance $d_{\mathrm{r},0}$ and path-gain exponent $\eta$, the received signal power at distance $d$ is usually expressed in dB as $P_{\mathrm{r}}=P_{\mathrm{r},0}+10\eta\mathrm{log}_{10}(d_{\mathrm{r},0}/d)$. Based on our experiments, we have chosen a similar model to show the relation between the energy of the slowly-varying signal and the the delay of the reflected paths.} and defined as
\begin{eqnarray}
G(\tau_{\mathrm{p},l})=G(\tau_{\mathrm{ref}})\left(\frac{\tau_{\mathrm{ref}}}{\tau_{\mathrm{p},l}}\right)^{\eta} \label{eq:PathGain}
\end{eqnarray}
for a reference path gain $G(\tau_{\mathrm{ref}})$ at a reference delay $\tau_{\mathrm{ref}}$ and a path-gain exponent $\eta$. The random variable $w(kT_{\mathrm{rep}})$ is of low frequency, zero mean and unit energy in time dimension (i.e., $\mathbb{E}\left\{\sum_{k}w^2(kT_{\mathrm{rep}})\right\}=N_{\mathrm{rep}}$) and has unknown statistical properties in the delay dimension. Furthermore, $w(kT_{\mathrm{rep}})$ is assumed to be independent from the noise $n(kT_{\mathrm{rep}},m/W)$. Finally, $V(m/W-\tau_{\mathrm{p},l})$ represents the influence from the transmitted pulse $p(m/W-\tau_{\mathrm{p},l})$ to the time-varying received signal part.

Note that in our setting, $L$, $\alpha_{l}$, and $\tau_{l}$ are not known to the localization
system, but assumed to be constant as a function of time, during $1\le k\le N_{\mathrm{rep}}$. Moreover, for the transmission between transmitter $i$ to receiver $j$, the human-body-induced delay of our interest is given as 
\begin{eqnarray}
\tau_{\mathrm{p},1}=\tau_{1}+\frac{||\mathbf{x}-\mathbf{x}_{i}||+||\mathbf{x}-\mathbf{x}_{j}||-||\mathbf{x}_{i}-\mathbf{x}_{j}||}{c},\label{eq:delayModel}
\end{eqnarray}
where $c$ is the speed of light. Since we can estimate $\tau_{1}$
using standard time-of-arrival estimation techniques\cite{Falsi}, and since we
know the positions of the anchors, and thus also $||\mathbf{x}_{i}-\mathbf{x}_{j}||$,
from (\ref{eq:delayModel}) it is straightforward to determine $\hat{d}_{i,j}$,
defined in (\ref{eq:range}).
Furthermore, we will abbreviate $r(kT_{\mathrm{rep}},m/W)$ as $r(k,m)$ with the
understanding that samples in the different dimensions (time, delay)
are taken at different rates. In this proof-of-concept study, we will not consider the statistics
of the measurement error $e_{i,j}$, or the impact of mobility and
corresponding tracking methods.

Our main objectives are to determine (i) if
the person is present
and (ii) on which ellipse the person is (i.e., what is $\tau_{\mathrm{p},1}$?).
{\subsubsection*{Comment} We should point out that our signal model assumes no other sources can create low-frequency disturbances to the UWB waveforms. In practice, slowly moving equipment or periodically moving stationary objects (e.g., a fan) may have similar effects as a human being on the UWB waveforms. Further study is required to incorporate these error sources.
\section{Detection and Device-Free Ranging}

\label{sec:Detection}

In this section, we develop a device-free person detection and a ranging
technique, based on the signal model in (\ref{eq:2DreceivedSignalModel}). We
will proceed as follows: we first consider a fixed delay $m/W$ and
determine a meaningful (though not necessarily sufficient or minimal)
statistic. Then, we combine these statistics over multiple delays,
allowing us to detect the presence of a person and to infer $\tau_{\mathrm{p},1}$.
An analysis of the final statistic is deferred to Section \ref{sec:Performance}.

\subsection{Statistic for Fixed Delay}

Considering a fixed delay $m/W$ in (\ref{eq:2DreceivedSignalModel}), we obtain
signals as shown in Fig.~\ref{fig:receivedWaveformFixedTauEmpty}. As the
mean of the signal does not convey relevant information for our purpose,
we will subtract it, thus leading to the following signal model
\begin{equation}
r_{m}(k)=\left\{ \begin{array}{ll}
n_{m}(k), & \quad\mathrm{no\, person\, affects\, delay\, m}\\
x_{m}(k)+n_{m}(k), & \quad\mathrm{person\, affects\, delay\, m},
\end{array}\right. \label{eq:signalModelFixedDelay}
\end{equation}
where we have moved the delay index $m$ to a subscript to emphasize
the dependence on the time dimension $k$. Here, $x_{m}(k)$ is a low-frequency
signal induced by the presence of the person and $n_{m}(k)$ is assumed to be white Gaussian noise with variance $N_{0}W/2$. In order to filter out high-frequency noise components from the low-frequency variations induced by the person, we decompose the received signal as
\begin{eqnarray}
r_{m}(k)=r_{m,\mathrm{L}}(k)+r_{m,\mathrm{H}}(k)\label{eq:receivedSignalLowHigh}
\end{eqnarray}
where $r_{m,\mathrm{L}}(k)$ and $r_{m,\mathrm{H}}(k)$ are the low-
and the high-frequency components, with fractional bandwidths (i.e., the bandwidth that is normalized to $1/T_{\mathrm{rep}}$) of $\beta$
and $1-\beta$, respectively, where $\beta\ll1$. In practice, $\beta$ depends upon how fast the movements of the person occur within the time window of $T_{\mathrm{rep}}N_{\mathrm{rep}}$. For instance, $\beta/T_{\mathrm{rep}}$ should be around 0.2--0.4 Hz for normal breathing.
Stacking the time samples yields the vectors $\mathbf{r}_{m}$, $\mathbf{x}_{m}$, and $\mathbf{r}_{m,\mathrm{L}}$. A likelihood ratio
test then yields a test statistic
\begin{align}
\frac{\exp\left(-\frac{1}{N_{0}W}\left\Vert \mathbf{r}_{m}-\mathbf{x}_{m}\right\Vert ^{2}\right)}{\exp\left(-\frac{1}{N_{0}W}\left\Vert \mathbf{r}_{m}\right\Vert ^{2}\right)}=\exp\left(\frac{2\mathbf{r}_{m}^{T}\mathbf{x}_{m}}{N_{0}W}-\frac{\left\Vert \mathbf{x}_{m}\right\Vert ^{2}}{N_{0}W}\right).
\end{align}
Since $\mathbf{x}_{m}$ is unknown, we treat it as a deterministic vector in the likelihood ratio test and replace it with an estimate.
We only exploit the low-pass nature of $\mathbf{x}_{m}$, and thus
set the estimate to $\hat{\mathbf{x}}_{m}=\mathbf{r}_{m,\mathrm{L}}$. We easily find
a final statistic, after normalization with $\beta N_{\mathrm{rep}}$ as\footnote{In practice, $N_{0}$ must be estimated as well. This can easily be
achieved using standard techniques \cite{Pauluzzi}. We will return to this in Section \ref{sec:Experimental-Results}.}
\begin{align}
y(m) & = \frac{\left\Vert \mathbf{r}_{m,\mathrm{L}}\right\Vert ^{2}}{N_{0}\beta WN_{\mathrm{rep}}}\label{eq:decisionStatisticFinalBintoBin}
\end{align}
where, in the absence of the person, $\mathbb{E}\{y(m)\}=1$.
\subsection{Statistic for Delay Window} \label{sec:statisticDelayWindow}

Up to now, we focused on a single delay value, however the person will affect a window of delays. Assuming that the person has an effect over the duration of the transmitted pulse
$T_{\mathrm{p}}$, we can aggregate the information over multiple delays by averaging the delay-specific statistic over a window around a trial delay $\tau$, which is assumed to be an integer multiple of $1/W$:
\begin{equation}
D(\tau)=\frac{1}{T_{\mathrm{p}}W}\sum_{m=\left(\tau-T_{\mathrm{p}}/2\right)W}^{\left(\tau+T_{\mathrm{p}}/2\right)W}y(m).\label{eq:decisionStatisticMultipleBin1}
\end{equation}
The presence of a person can thus be determined by comparing $D(\tau)$
to a threshold. Hence, the person is detected as
\begin{equation}
\begin{cases}
D(\tau)\le\gamma & \mathrm{no\, person\, present}\\
D(\tau)>\gamma & \mathrm{person\, present},
\end{cases}\label{eq:decisionRule}
\end{equation}
where $\gamma$ is a threshold. The selection of $\gamma$
depends on the desired performance trade-off, and will be treated
in Section~\ref{sec:Performance}.

In addition to the presence of the person, $D(\tau)$ also conveys information about the delay of the human body induced reflections, from which we can obtain an estimate of $\tau_{\mathrm{p},1}$. Here, we introduce three practical estimation criteria:
\subsubsection{Line Search}This criterion is based on the selection of the largest sample in $D(\tau)$, and specifically given by
\begin{equation}
\hat{\tau}_{\mathrm{p},1}=\arg\max_{\tau \in [0,T_{\mathrm{delay}}]}D(\tau)\label{eq:decisionStatisticMultipleBin2}.
\end{equation}
\subsubsection{Threshold Crossing}The maximum value of $D(\tau)$ needs not always correspond to $\tau_{\mathrm{p},1}$ but may come from a later arriving indirect reflected path $\tau_{\mathrm{p},l}$ with $l=2,\ldots,M$. A detailed discussion of this phenomenon is left to Section~\ref{sec:Experimental-Results}. To provide robustness against these failures, we introduce another criterion where we estimate $\tau_{\mathrm{p},1}$ based on the first threshold crossing:
\begin{equation}
\hat{\tau}_{\mathrm{p},1}=\min \left\{ {\tau}: D(\tau)> \tilde{\gamma} \right\}.\label{eq:thresholdBasedRanging}
\end{equation}
$\tilde{\gamma}$ is also a threshold and can be different than $\gamma$, depending on the desired performance and false alarm criteria.
\subsubsection{Maximum Rise Search}Although the previous approach provides additional robustness, its performance is dependent on the threshold value, which may be hard to determine in some cases. We can also exploit the shape of $D(\tau)$ in which we search for the maximum increase and thereby remove the requirement for the threshold determination. Specifically, the maximum rise search criterion is given by
\begin{equation}
\hat{\tau}_{\mathrm{p},1}=\arg\max_{\tau \in [0,T_{\mathrm{delay}}]}\frac{D(\tau+T_{\mathrm{win}}/2)-D(\tau-T_{\mathrm{win}}/2)}{T_{\mathrm{win}}}\label{eq:maximumRiseSearch}
\end{equation}
where $T_{\mathrm{win}}$ is an arbitrary window (whose length we will set to $T_{\mathrm{win}}=T_{\mathrm{p}}$, neglecting the pulse distortion in the channel).

\section{Performance Analysis}

\label{sec:Performance}In this section we determine the false alarm and missed detection probabilities of the detector proposed
in (\ref{eq:decisionRule}), for an estimate $\tau^*$, which can be generated from any of the three above-mentioned criteria.

\subsection{Probability of False Alarm}

In the absence of the person, substitution of (\ref{eq:decisionStatisticFinalBintoBin})
into (\ref{eq:decisionStatisticMultipleBin1}), and accounting for
the fact that $r_{m,\mathrm{L}}(k)=n_{m,\mathrm{L}}(k)$ is
low-frequency noise, yields
\begin{equation}
D(\tau^{*})=\frac{1}{T_{\mathrm{p}}N_{0}\beta W^{2}N_{\mathrm{rep}}}\sum_{m=\left(\tau^{*}-T_{\mathrm{p}}/2\right)W}^{\left(\tau^{*}+T_{\mathrm{p}}/2\right)W}\left\Vert \mathbf{n}_{m,\mathrm{L}}\right\Vert ^{2}.\label{eq:PFA1}
\end{equation}
For sufficient number of time samples $N_{\mathrm{rep}}$, we can invoke the
central limit theorem (CLT), and approximate $D(\tau^{*})\sim\mathcal{N}(\mu_{\mathrm{f}},\sigma_{\mathrm{f}}^{2})$,
where
\begin{equation}
\mu_{\mathrm{f}}=1
\end{equation}
and $\sigma_{\mathrm{f}}^{2}$ is derived in
Appendix \ref{subsec:Apendix1}, and found to be
\begin{align}
\sigma_{\mathrm{f}}^{2} & =  \frac{2}{T_{\mathrm{p}}\beta WN_{\mathrm{rep}}}.
\end{align}
Hence, the probability of false alarm $P_{\mathrm{FA}}$ is given
as
\begin{align}
P_{\mathrm{FA}} & = \mathrm{Pr}\left\{ D(\tau^{*})>\gamma\,|\,\mathrm{no\, person\, present}\right\} \nonumber \\
 & =  Q\left(\frac{\gamma-\mu_{\mathrm{f}}}{\sigma_{\mathrm{f}}}\right),
\end{align}
 where $Q(.)$ denotes the Q function.

\subsection{Probability of Missed Detection}

In the presence of the person, substitution of (\ref{eq:decisionStatisticFinalBintoBin})
into (\ref{eq:decisionStatisticMultipleBin1}), and accounting for
the fact that $r_{m,\mathrm{L}}(k)=x_{m}(k)+n_{m,\mathrm{L}}(k)$,
yields
\begin{equation}
D(\tau^{*})=\frac{1}{T_{\mathrm{p}}N_{0}\beta W^{2}N_{\mathrm{rep}}}\sum_{m=\left(\tau^{*}-T_{\mathrm{p}}/2\right)W}^{\left(\tau^{*}+T_{\mathrm{p}}/2\right)W}\left\Vert \mathbf{x}_{m}+\mathbf{n}_{m,\mathrm{L}}\right\Vert ^{2}.\label{eq:PMD1}
\end{equation}
From the right-hand side of (\ref{eq:fixedTimeVarying}), $x_{m}(k)$ is written for the direct reflected path as
\begin{eqnarray}
x_{m}(k) =  f(\tau_{\mathrm{p},1})w(k)V(m/W-\tau_{\mathrm{p},1}).\label{eq:PMD3}
\end{eqnarray}
Substitution of (\ref{eq:PMD3}) into (\ref{eq:PMD1}), expanding the square, and
considering that $\sum_{m}V^2(m/W-\tau_{\mathrm{p},1})=W$ yields
\begin{align}
D(\tau^{*}) & =\frac{1}{T_{\mathrm{p}}N_{0}\beta W^{2}N_{\mathrm{rep}}}\sum_{m=\left(\tau^{*}-T_{\mathrm{p}}/2\right)W}^{\left(\tau^{*}+T_{\mathrm{p}}/2\right)W}\Bigl(\left\Vert \mathbf{x}_{m}\right\Vert ^{2}+\left\Vert \mathbf{n}_{m,\mathrm{L}}\right\Vert ^{2}\nonumber\\
&+2\mathbf{x}_{m}^{T}\mathbf{n}_{m,\mathrm{L}}\Bigr) \nonumber\\
 & =\frac{G(\tau_{\mathrm{p},1})E_s}{T_{\mathrm{p}}N_{0}\beta W^{2}N_{\mathrm{rep}}}\sum_{m=\left(\tau^{*}-T_{\mathrm{p}}/2\right)W}^{\left(\tau^{*}+T_{\mathrm{p}}/2\right)W}V^{2}(m/W-\tau_{\mathrm{p},1})\left\Vert \mathbf{w}\right\Vert ^{2} \nonumber \\
& +\frac{1}{T_{\mathrm{p}}N_{0}\beta W^{2}N_{\mathrm{rep}}}\sum_{m=\left(\tau^{*}-T_{\mathrm{p}}/2\right)W}^{\left(\tau^{*}+T_{\mathrm{p}}/2\right)W}\left\Vert \mathbf{n}_{m,\mathrm{L}}\right\Vert ^{2} \nonumber\\
&+\frac{2}{T_{\mathrm{p}}N_{0}\beta W^{2}N_{\mathrm{rep}}}\sum_{m=\left(\tau^{*}-T_{\mathrm{p}}/2\right)W}^{\left(\tau^{*}+T_{\mathrm{p}}/2\right)W}\mathbf{x}_{m}^{T}\mathbf{n}_{m,\mathrm{L}}.
\end{align}
Invoking again the CLT for sufficiently large $N_{\mathrm{rep}}$, $D(\tau^{*})\sim\mathcal{N}(\mu_{\mathrm{d}},\sigma_{\mathrm{d}}^{2})$,
where
\begin{align}
\mu_{\mathrm{d}} =  1+\frac{G(\tau_{\mathrm{p},1})E_{s}}{T_{\mathrm{p}}N_{0}\beta W}.\label{eq:mu_{d}}
\end{align}
 and (see Appendix \ref{subsec:Apendix2} for details)
\begin{eqnarray*}\label{eq:sigma_{d}}
\sigma_{\mathrm{d}}^{2}  =  \frac{2}{T_{\mathrm{p}}\beta WN_{\mathrm{rep}}}+ \frac{4G(\tau_{\mathrm{p},1})E_{s}}{T_{\mathrm{p}}^2N_{0}\beta^2W^2N_{\mathrm{rep}}}.
\end{eqnarray*}
Finally, the probability of missed detection $P_{\mathrm{MD}}$ is
obtained as
\begin{align}
P_{\mathrm{MD}} & =  \mathrm{Pr}\left\{ D(\tau^{*})\le\gamma\,|\,\mathrm{person\, present}\right\} \nonumber \\
 & =  1-Q\left(\frac{\gamma-\mu_{\mathrm{d}}}{\sigma_{\mathrm{d}}}\right).
\end{align}

\section{Performance Evaluation and Discussion} \label{sec:Performance Evaluation}

In this section, we quantify the performance of the detection and the device-free ranging technique from Section~\ref{sec:Detection}. We will first consider the detection and provide numerical results for the relevant performance measures. Then, we will show experimental results for both detection and the device-free ranging. Finally, the localization performance will be evaluated.

\subsection{Detection and Device-Free Ranging: Simulation Results}\label{sec:Numerical-Results}
\subsubsection*{Simulation Setup}
Unless otherwise stated, we consider the $7$-th derivative Gaussian pulse as a transmitted signal with a duration of $T_{\mathrm{p}}=1.4$~ns and set the receiver sampling rate to $61$~ps, resulting in $23$ delay samples over the transmitted pulse duration. We collect a varying number of repetitions with a period of $T_\mathrm{rep}= 0.2~\mathrm{s}$. Without loss of generality, the variation due to the person is considered to have a sinusoidal shape\cite{Venkatesh,Gezici2} with a normalized frequency of $0.04$~cycles/sample (i.e., an absolute frequency of 0.2 Hz). The fractional bandwidth of the filter, applied over time, is set to $\beta=0.1$ (i.e., an absolute bandwidth of 0.5 Hz). Hence, it is fully possible to capture the temporal variations induced by the person. Based on our measurement data\footnote{We first define a reference received signal as the first $T_{\mathrm{p}}W$ samples of the received waveform after the leading edge at a given transmitter and the receiver distance. For the same distance, we calculate the total energy of the human-body induced delay samples (again over the length of $T_{\mathrm{p}}W$) after removing the mean from the signal. Finally, we divide by the energy of the reference signal. While extracting these parameter values, we only consider the cases when the person is not blocking the direct line-of-sight (i.e we exclude the shadowing effect induced by the person).}, we set the model parameters in (\ref{eq:PathGain}) as $G(\tau_{\mathrm{ref}})=3.6\times10^{-3}$, $\tau_{\mathrm{ref}}=17.8$~ns and $\eta=5.5$. Finally, the device-free range is denoted as $d_{0}=c\tau_{\mathrm{p},1}$, and the received signal-to-noise ratio (SNR) is defined as $\mathrm{SNR}=E_{s}/N_{0}$, where we have scaled the signal with $1/\alpha_1$.

\subsubsection*{Results and Discussion}

\begin{figure}[!t]
\begin{center}
\psfrag{xlabel}{\footnotesize{$\gamma$}}
\psfrag{ylabel}{\footnotesize{$P_{\mathrm{FA}}$}}
\psfrag{NREP1000XXXXX}{\footnotesize{$N_{\mathrm{rep}}=100$}} 
\psfrag{NREP500XXXXX}{\footnotesize{$N_{\mathrm{rep}}=200$}}
\psfrag{NREP200XXXXX}{\footnotesize{$N_{\mathrm{rep}}=500$ }}
\psfrag{NREP100XXXXX}{\footnotesize{$N_{\mathrm{rep}}=1000$}}
\psfrag{Analytical}{\footnotesize{Analytical}}
\includegraphics[width=1\columnwidth]{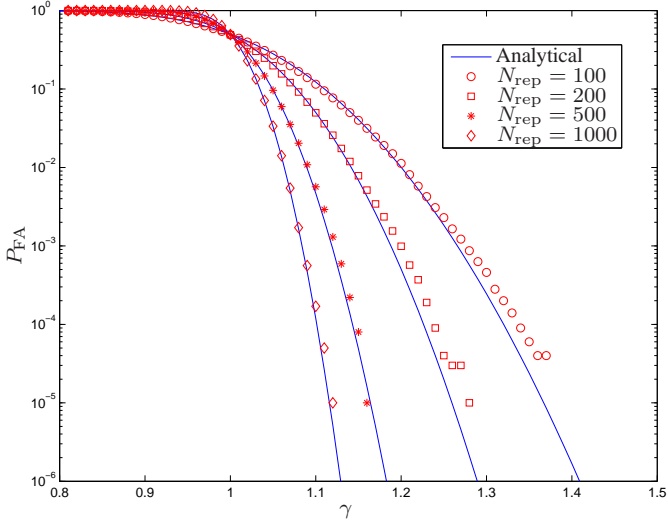}
\end{center}
\caption{False alarm probability versus threshold under different number of samples over time.}
\label{fig:FalseAlarmVsThreshold}
\end{figure}

\begin{figure}[!t]
\begin{center}
\hskip -0.8cm
\psfrag{xlabel}{\footnotesize{$\gamma$}}
\psfrag{ylabel}{\footnotesize{$P_{\mathrm{MD}}$}}
\psfrag{NREP100ANALYTICALXXX}{\footnotesize{$N_{\mathrm{rep}}=100$ (Analytical)}}
\psfrag{NREP1000ANALYTICALXXX}{\footnotesize{$N_{\mathrm{rep}}=1000$ (Analytical)}}
\psfrag{NREP100SIMULATIONXXX}{\footnotesize{$N_{\mathrm{rep}}=100$ (Simulation)}}
\psfrag{NREP1000SIMULATIONXXX}{\footnotesize{$N_{\mathrm{rep}}=1000$ (Simulation)}}
\psfrag{LOWSNRXX}{\footnotesize{Low SNR}}
\psfrag{HIGHSNRXX}{\footnotesize{High SNR}}
\includegraphics[width=1.08\columnwidth]{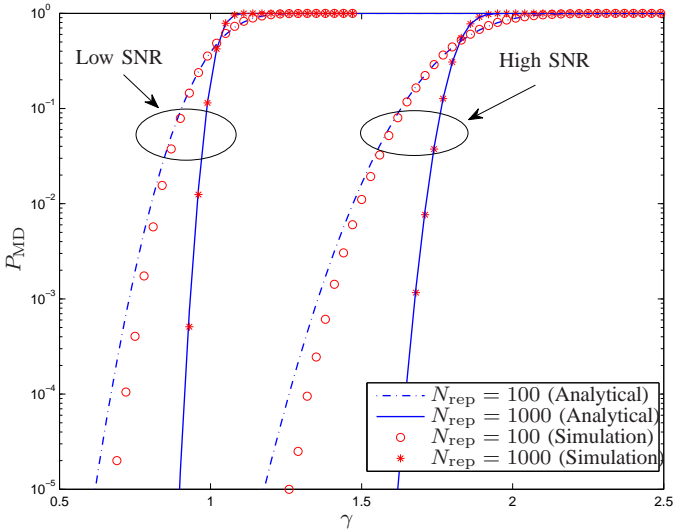}
\end{center}
\caption{Missed detection probability versus threshold under different number of samples over time at low (15~dB) and high (30~dB) SNR and $d_0=6$~m.}
\label{fig:MissedDetectionVsThreshold}
\end{figure}

We first validate false alarm and missed detection probabilities derived in the previous section. The validation was based on simulating the system described above and determining the false alarm and missed detection probabilities through Monte Carlo estimation.
Fig.~\ref{fig:FalseAlarmVsThreshold} shows the false alarm probability as a function of the threshold $\gamma$ for different values of $N_{\mathrm{rep}}$. Note that this performance does not depend on the SNR. We observe good agreement between the simulations and the predicted performance, especially for larger values of $N_{\mathrm{rep}}$, as involving the CLT becomes more valid when more observations are collected. Fig.~\ref{fig:MissedDetectionVsThreshold} shows the missed detection probability as a function of the threshold $\gamma$, for two different SNR values, and different values of $N_{\mathrm{rep}}$. We again observe good agreement for large $N_{\mathrm{rep}}$, for both low and high SNR. When $N_{\mathrm{rep}}$ is low, the predicted performance is worse than the simulated performance. This is due to the fact that the Gaussian approximation does not completely fit to the distribution of (\ref{eq:PMD1}). The mismatch occurs in left-side tail, where the Gaussian approximation underestimates the true probability density. Note that, in general, $N_{\mathrm{rep}}$ is dependent on both total observation duration over time window and the measurement update rate ($1/T_{\mathrm{rep}}$). Therefore, the performance can be described well with analytical results for  high update rates over shorter time windows (sufficiently large to capture slow variations) or slower update rates (again sufficiently fast to capture slow variations) over longer time windows. In the following part, we will only show the analytical results assuming that update rate is sufficiently high to collect $N_{\mathrm{rep}}=1000$ samples over the time window.

\begin{figure}[!t]
\begin{center}
\psfrag{xlabel}{\footnotesize{SNR [dB]}}
\psfrag{ylabel}{\footnotesize{$P_{\mathrm{MD}}$}}
\psfrag{PFA01XXXXXXXX}{\footnotesize{$P_{\mathrm{FA}}=10^{-1}$}}
\psfrag{PFA001XXXXXXXX}{\footnotesize{$P_{\mathrm{FA}}=10^{-2}$ }}
\psfrag{PFA0001XXXXXXXX}{\footnotesize{$P_{\mathrm{FA}}=10^{-3}$}}
\psfrag{PFA00001XXXXXXXX}{\footnotesize{$P_{\mathrm{FA}}=10^{-4}$}}
\includegraphics[width=1\columnwidth]{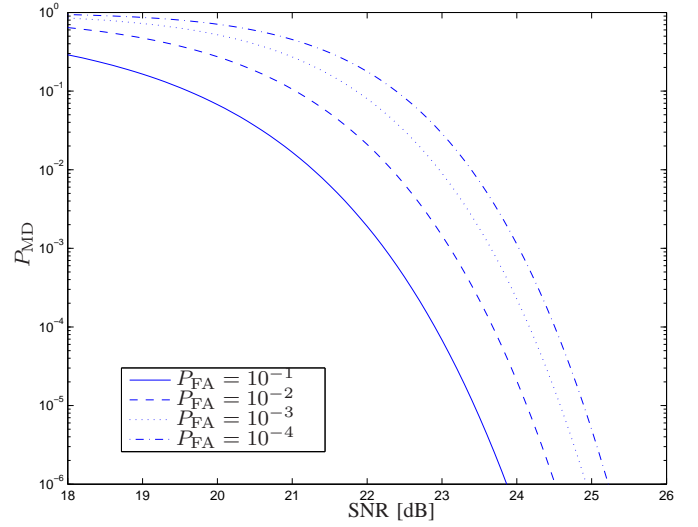}
\end{center}
\caption{Missed detection probability versus SNR under different false alarm probabilities ($d_0=6$~m, $N_{\mathrm{rep}}=1000$).}
\label{fig:MissedDetectionVsSNR}
\end{figure}

Figs.~\ref{fig:MissedDetectionVsSNR} and Fig.~\ref{fig:MissedDetectionVsReflectionDistance} show the dependence of the missed detection probability on SNR and the distance, respectively, for varying values of the false alarm probability $P_{\mathrm{FA}}$. Fig.~\ref{fig:MissedDetectionVsSNR} illustrates that for a given device-free range $d_{0}$ and $P_{\mathrm{FA}}$, we need about 1~dB of SNR improvement for every order of magnitude reduction in $P_{\mathrm{MD}}$ (e.g., around $3$~dB to get from $P_{\mathrm{MD}}=10^{-1}$ to $P_{\mathrm{MD}}=10^{-4}$).
Furthermore, the performance of the system also depends on the position of the person with respect to the transmitter and the receiver distance, as we observed, the slow-variation of the signal attenuates by the increase in the reflected-path length. Fig.~\ref{fig:MissedDetectionVsReflectionDistance} shows that, considering $P_{\mathrm{MD}}=10^{-3}$, it is possible to detect the person up to $10$~m device-free range at high SNR (e.g., $35$~dB). Note that the device-free range is always larger than the arrival time of the signal ($d_0>d_1$). Therefore, although results show that at higher SNRs the device-free range increases, the SNR decreases with the distance between the transmitter and the receiver due to the path loss. Thus, in practice the device-free range is limited by the transmitter and the receiver distance.
\begin{figure}[!t]
\begin{center}
\psfrag{xlabel}{\footnotesize{$d_{0}$ [m]}}
\psfrag{ylabel}{\footnotesize{$P_{\mathrm{MD}}$}}
\psfrag{SNR20XXXXXXXX}{\footnotesize{$\mathrm{SNR}=20$ dB}}
\psfrag{SNR25XXXXXXXX}{\footnotesize{$\mathrm{SNR}=25$ dB}}
\psfrag{SNR30XXXXXXXX}{\footnotesize{$\mathrm{SNR}=30$ dB}}
\psfrag{SNR35XXXXXXXX}{\footnotesize{$\mathrm{SNR}=35$ dB}}
\includegraphics[width=1\columnwidth]{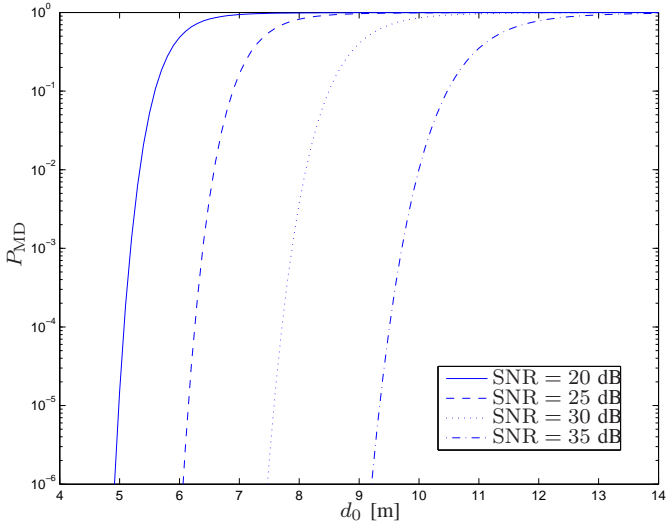}
\end{center}
\caption{Missed detection probability versus device-free range under different channel conditions ($P_{\mathrm{FA}}=10^{-3}$, $N_{\mathrm{rep}}=1000$).}
\label{fig:MissedDetectionVsReflectionDistance}
\end{figure}

\subsection{Detection and Device-Free Ranging: Experimental Results}\label{sec:Experimental-Results}

\begin{figure}[!t]
\vskip -0.5cm
\begin{center}
\hskip -0.6cm
\psfrag{xlabel}{\footnotesize{x [m]}}
\psfrag{ylabel}{\footnotesize{y [m]}}
\psfrag{TxXX}{\footnotesize{Tx}}
\psfrag{RxXX}{\footnotesize{Rx}}
\psfrag{H1X}{\footnotesize{H1}}
\psfrag{H2X}{\footnotesize{H2}}
\psfrag{H3X}{\footnotesize{H3}}
\psfrag{H4X}{\footnotesize{H4}}
\psfrag{H5X}{\footnotesize{H5}}
\psfrag{H6X}{\footnotesize{H6}}
\psfrag{H7X}{\footnotesize{H7}}
\psfrag{H8X}{\footnotesize{H8}}
\psfrag{H9X}{\footnotesize{H9}}
\psfrag{H10X}{\footnotesize{H10}}
\psfrag{H11X}{\footnotesize{H11}}
\psfrag{H12X}{\footnotesize{H12}}
\psfrag{H13X}{\footnotesize{H13}}
\includegraphics[width=1.06\columnwidth]{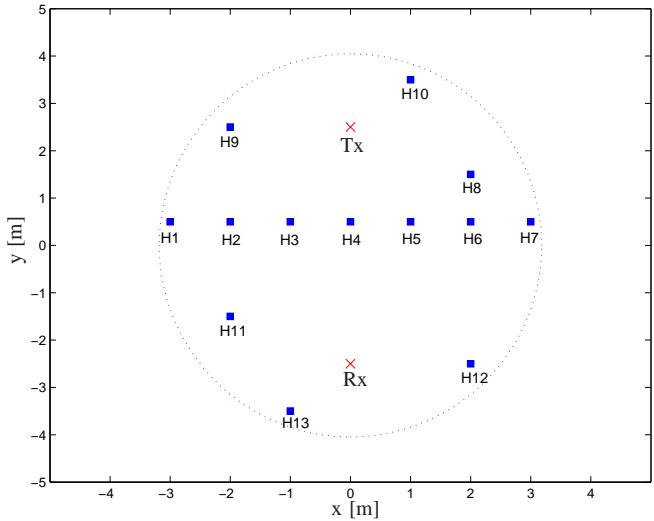}
\end{center}
\caption{The overview of measurement scenarios: the transmitter (Tx), the receiver (Rx) (red cross) and the subject positions (blue rectangular) and device-free detection range (dotted ellipse) limited by the current radios. }
\label{fig:measurementScenarios}
\end{figure}

\begin{figure}[!t]
\begin{center}
\includegraphics[width=0.9\columnwidth]{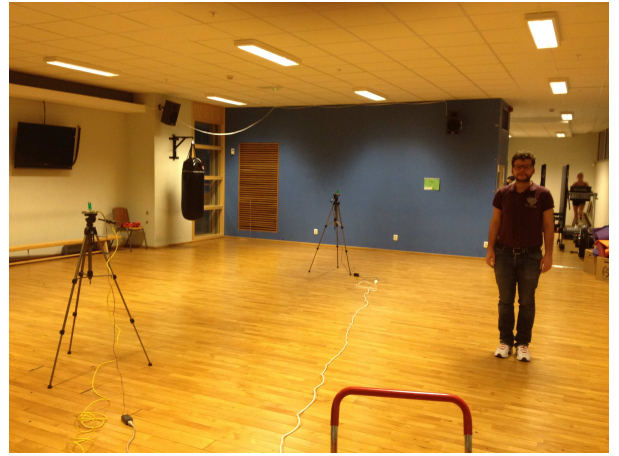}
\end{center}
\caption{UWB setup for device-free localization measurements, the person is standing at H1 with a perpendicular body orientation.}
\label{fig:measurementPicture}
\end{figure}

\subsubsection*{Experiment Setup}
We performed experiments in a fitness room at the Chalmers University of Technology building with two identical, commercially available and FCC-compliant UWB radios (Time Domain P400 radios\footnote{See http://www.timedomain.com/}). The radios are capable of performing communications and ranging using UWB signaling within the frequency range of 3--5.5~GHz. They were equipped with identical antennas that are omni-directional within the operating frequency range. While these radios are primarily designed for two-way time-of-arrival ranging, they are capable of capturing a waveform. At the receiving radio, the waveforms are sampled with a sampling period of $61$~ps over a window of approximately 10 ns, limiting the maximum captured multi-path length to be less than approximately $3$~m. With the current hardware release, beyond the 10 ns window, captured waveforms are no longer stable and processing across time is not meaningful. The transmitted pulse duration was measured with high-frequency oscilloscope and found to be approximately $1.4$~ns.

Measurements were performed with fixed transmitter (Tx) and receiver (Rx) positions, while the person was standing at 13 different points as shown in Fig~\ref{fig:measurementScenarios}. For each position, the person stood in two orthogonal  orientations, so that the body is either perpendicular or parallel to the line-of-sight between radios. During the experiments, the radios were mounted on tripods at a height of $1.23$~m above the ground and were connected via ethernet cables. Measurements were taken using the graphical user interface at a rate of $5$~measurements/second. The waveforms were collected over 20 seconds, allowing us to obtain $100$ snapshots over time for each scenario. During the experiments, care was taken to keep the environment static. Therefore only the subject was standing within the device-free detection range limited by the radios. The operator, who was controlling the experiments, and all the other people were outside the range. A typical measurement scenario is shown in Fig.~\ref{fig:measurementPicture}. During the off-line postprocessing, fine alignment around the leading edge point was performed by cross-correlation to make sure that all the waveforms are aligned over the time window. The starting instant of the waveform (the zero delay instant) is defined based on the leading edge detection point, provided by the radio. Similar to the simulations settings, we applied the low-pass filter with a fractional bandwidth of $\beta=0.1$. We chose the detection threshold as $\gamma=1.38$ to obtain a false alarm probability of $P_{\mathrm{FA}} \approx 10^{-5}$ (see also Fig.~\ref{fig:FalseAlarmVsThreshold}).

\emph{Note on background noise and timing jitter:} The radios are subject to a number of hardware impairments, in particular the timing jitter. In general, the timing jitter causes deviation of the transmitted pulses from reception at integer multiples of sampling time\cite{Guvenc_Jitter}. At a certain delay sample, the signal amplitude varies over time proportional to the time rate of change of the received waveform (i.e., the derivative of received waveform to time). This variation will be greater for higher amplitudes, since the change in the amplitude will be greater, leading to a higher impact of the timing jitter for these samples. This resulted in different amounts of background noise for delay with high signal values, compared to delays with low signal values. For that reason, we estimate $N_0$ on a delay by delay basis. For a given delay, $m$, we estimate the noise power $N_{0,m}$ as
\begin{eqnarray}\label{eq:noiseEstimation}
\hat{N}_{0,m}=\frac{1}{W(1-\beta)N_{\mathrm{rep}}}\sum_{k=1}^{N_{\mathrm{rep}}}r^{2}_{m,\mathrm{H}}(k).
\end{eqnarray}
where $r_{m,\mathrm{H}}(k)=r_{m}(k)-r_{m,\mathrm{L}}(k)$, is the part of the signal that contains only noise, irrespective of the presence of a person. The estimate was then substituted in (\ref{eq:decisionStatisticFinalBintoBin}) to allow detection and ranging in the presence of hardware imperfections.

\subsubsection*{Experiment Results and Discussion}\label{sec:DFRExperimentalResults}

Experimental results for the device-free ranging are tabulated in Table~\ref{table:ExperimentalResults}, showing, for each of the measurement locations, the true device-free range, the device-free range error (calculated as the difference between the estimated and the true device-free ranges, and averaged over the body orientations) for three ranging criteria introduced in Section~\ref{sec:statisticDelayWindow}. Our results show that the proposed method was able to detect the presence of the person in all cases. Furthermore, when we employ the line search approach, the device-free range error is found to be always positive including different orientations and less than $60$~cm for all the cases, except the case where the person blocks the direct line-of-sight (case H4). A deeper investigation revealed that temporal variations are not only observed at a single delay window, but at multiple delay windows of the received waveform. This effect is not dominant in other cases and may be due to the fact that the signal is reflected by another reflection source, after being reflected by the person. We also observe that the estimated noise power may be lower for samples related to these indirect reflections, resulting in higher values for the decision statistic (this is a side-effect of the way we dealt with timing jitter). On the other hand, when we employ threshold crossing approach, we obtain a better ranging performance as shown in Table~\ref{table:ExperimentalResults} for position H4. In this case, we kept the threshold $\tilde{\gamma}$ $=$ $\gamma$ $=$ $1.38$. Although the ranging performance improves for H4, the threshold crossing criterion results in early detections in some other positions. This is again mainly due to the variation of the estimated noise power across delay samples, which has lower values for early delay samples, resulting in higher values for the delay statistic. Increasing the threshold may lead to better results, though, in general, the determination of the optimum value might be a difficult task. Finally, as the maximum rise search criterion removes the requirement of the threshold determination, it outperforms the threshold crossing and gives a better ranging performance compared to the line search for H4.

\begin{table}[!t]
\renewcommand{\arraystretch}{0.9}
\caption{Experimental Results for Device-Free Ranging}
\label{table:ExperimentalResults}
\centering
\begin{tabular}{|c||c||rrr|}
\hline
  &  & \multicolumn{3}{|c|}{\bfseries Error [m]} \\
\bfseries Position & \bfseries Range [m] & Line & Threshold & Maximum  \\
 & & search & crossing & rise search \\ \hline \hline
 H1 & $7.85$ & $0.2$ & $-1.22$ & $0.01$  \\
 \hline
  H2 & $6.43$ & $0.4$ & $-0.55$ & $0.09$  \\
   \hline
  H3 & $5.4$ & $0.55$ & $0.15$ & $0.17$\\
  \hline
  H4 & $5$ &$1.4$ & $0.42$ & $0.81$  \\
  \hline
  H5 & $5.4$ & $0.6$ & $0.25$ & $ 0.29$  \\
   \hline
  H6 & $6.43$ & $0.38$ & $0.01$  & $0.05$  \\
   \hline
  H7 & $7.85$ & $0.19$ & $-1.29$ & $-0.12$  \\
    \hline
  H8 &$6.7$ & $0.27$  & $-0.85$ & $-0.14$  \\
    \hline
  H9 & $7.39$ & $0.27$  & $-1.07$ & $-0.03$ \\
    \hline
 H10 & $7.5$ & $0.07$ & $-1.24$ & $-0.31$  \\
    \hline
  H11 & $6.7$ &$0.44$ & $0.06$  & $0.02$ \\
   \hline
  H12 & $7.39$ & $0.32$ & $-0.13$  & $ -0.04$ \\
   \hline
  H13 & $7.5$ & $0.46$ & $-2.07$ & $0.17$ \\ \hline
\end{tabular}
\end{table}

Note that in our current work, we did not investigate the effect of other moving objects on the detection and the ranging performance. The detection method might give false alarms because of the changes in signal due to other autonomous systems, hence reducing the applicability of the system in some environments. Such changes in signal might depend on, for instance, the main material of the object and its shape and/or the strength or character of the movement, and should be considered by inspecting the dynamic nature of the environment when it is not populated (e.g., an industrial environment can be highly dynamic due to the machinery, whereas office environments are, most of the time, less dynamic). Effects due to the other non-stationary objects are left for future work.

\subsection{Device-Free Localization: Experiment Results}\label{sec:Numerical-Results-Loc}

While our current analysis did not include a statistical evaluation of the device-free range error, we here present indicative results of the localization capabilities of the proposed system. Experiments were performed in a room, located on the top floor of the Carr\'{e} building in University of Twente. The room was partly furnished with tables and chairs, and also has thick metallic pipes for ventilation and water supply. We deployed four anchor nodes, connected to a computer via network cables and a switch, and arranged in a square at positions $\mathbf{x}_{1}=(-5,0)$, $\mathbf{x}_{2}=(0,0)$, $\mathbf{x}_{3}=(-5,-5)$, $\mathbf{x}_{4}=(0,-5)$. A person stood on 24 different positions in a grid within a 3 meter by 5 meter area. For each position, we took six anchor-to-anchor measurements. Between each anchor pair, measurements were performed over 20~seconds with a rate of $50$~measurements/second (i.e., 1000 snapshots).

In Fig.~\ref{fig:locTC}, localization results are shown when we apply the threshold crossing criterion for device-free range estimates. The threshold is chosen as $\tilde{\gamma}=1.1$, which corresponds to $P_{\mathrm{FA}}$ of $10^{-5}$ for $N_{\mathrm{rep}}=1000$ based on the numerical results from Fig.~\ref{fig:FalseAlarmVsThreshold}. We get error values ranging from $0.12$~m (measurement point 18) to $1.8$~m (measurement point 3). 
The root mean square (RMS) localization error, computed as $\sqrt{1/N_{\mathrm{pos}}\sum_{i=1}^{N_{\mathrm{pos}}}||\mathbf{\widetilde{x}}_{\mathrm{p},i}-\mathbf{x}_{\mathrm{p},i}||^{2}}$ where $N_{\mathrm{pos}}$ denotes the number of standing positions of the person (i.e., 24 in this case) is $1$~m. When employing the maximum rise search and line search methods, the RMS localization error increases to $1.33$~m and $1.6$~m, respectively. Higher RMS localization errors can be explained similarly to our observations from Section~\ref{sec:DFRExperimentalResults}. In general, significant negative errors in device-free range estimates with threshold crossing criterion (i.e., when the person is close to the direct-LOS between devices), and significant positive errors with line search and maximum rise search criteria (i.e., when the person is away from the direct-LOS), leading to increased errors in the final position estimates.

\begin{figure}[!t]
\begin{center}
\psfrag{xlabel}{\footnotesize{x [m]}}
\psfrag{ylabel}{\footnotesize{y [m]}}
\includegraphics[width=1\columnwidth]{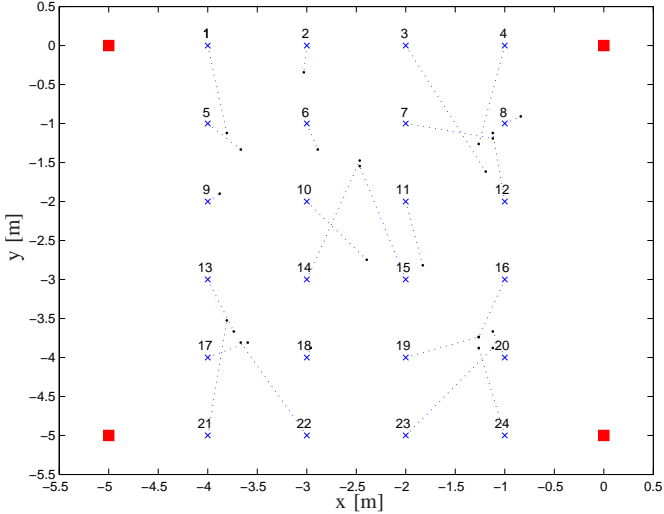}
\end{center}
\caption{Experimental results for localization performance, obtained when the person stands on 24 different positions in an indoor environment with four anchor nodes (marked as red squares). Threshold crossing criterion is applied to obtain the device-free range estimates.}
\label{fig:locTC}
\end{figure}

\section{Conclusion and Future Work}\label{sec:Conclusion}

In this paper, we developed a novel indoor UWB device-free person detection and ranging technique that does not require any knowledge about the environment. Our method relies solely on exploiting the temporal variations in the received signal induced by the presence of the person. We observed that the signal evolves slowly over time, even when the person is standing still. Based on this, we developed a signal model. Using this model, we formed a decision statistic to detect the person and estimate the delay of the human-body reflected signal from which we obtained the device-free ranges based on the three different criteria. We analyzed the detection probability and validate with simulation results. We also demonstrated the performance with experiments, which revealed that our method was able to detect the person and (for most of the cases) resulted in satisfactory device-free ranging performance. However, outliers in range estimates degrade the localization performance of the system which we also showed in an experimental environment. Future avenues of research include the development of ranging error models, locating multiple people, tracking of moving people and investigating the effect of other non-stationary objects in the environment.

\section*{Acknowledgement}Authors wish to thank Alan Petroff and Brandon Dewberry from TimeDomain Corp. for their helpful support with the measurement equipment; Pontus Johannisson, Martin Sj\"{o}din, Gabriel E.~Garcia, Pinar O\u{g}uz Ekim, Christopher Lindberg, and Philipp M\"{u}ller for their help with measurements and Kasra Haghighi for helpful discussions.

\appendix

\subsection{Derivation of Statistics for False Alarm Probability\label{subsec:Apendix1}}

We recall that in the absence of a person
\begin{eqnarray}
D(\tau^{*})=\frac{1}{T_{\mathrm{p}}N_{0}\beta W^{2}N_{\mathrm{rep}}}\sum_{m=\left(\tau^{*}-T_{\mathrm{p}}/2\right)W}^{\left(\tau^{*}+T_{\mathrm{p}}/2\right)W}\left\Vert \mathbf{n}_{m,\mathrm{L}}\right\Vert ^{2},
\end{eqnarray}
where $n_{m,\mathrm{L}}(k)$ are Gaussian noise samples obtained after low-pass filtering and with
variance $N_{0}W\beta$.
We first calculate
\begin{align}
\mathbb{E}\left\{ D^{2}(\tau^{*})\right\} & = \frac{1}{\left(T_{\mathrm{p}}N_{0}\beta W^{2}N_{\mathrm{rep}}\right)^{2}}\sum_{m,m'}\mathbb{E}\left\{ \left\Vert \mathbf{n}_{m,\mathrm{L}}\right\Vert ^{2}\left\Vert \mathbf{n}_{m',\mathrm{L}}\right\Vert ^{2}\right\}, \nonumber \\
& =  \frac{1}{T_{\mathrm{p}}^{2}W^{2}}\sum_{m,m'}\psi(m,m').
\end{align}
where the summation goes over interval of length $T_{\mathrm{p}}W$ (in two dimensions) and where we have introduced
\begin{align}
\psi(m,m')=\frac{1}{N_{0}^{2}\beta^{2}W^{2}N_{\mathrm{rep}}^{2}}\sum_{k,k'}\mathbb{E}\left\{ n_{m,\mathrm{L}}^{2}(k)n_{m',\mathrm{L}}^{2}(k')\right\}.
\end{align}
When $m\neq m'$, we find that
\begin{align}
\psi(m,m')  & =\frac{1}{N_{0}^{2}\beta^{2}W^{2}N_{\mathrm{rep}}^{2}}\sum_{k,k'}\mathbb{E}\left\{ n_{m,\mathrm{L}}^{2}(k)\right\}\mathbb{E}\left\{n_{m',\mathrm{L}}^{2}(k')\right\} \nonumber \\
 & =\frac{\left(N_{0}\beta WN_{\mathrm{rep}}\right)^{2}}{N_{0}^{2}\beta^{2}W^{2}N_{\mathrm{rep}}^{2}} \nonumber \\
 & = 1.
\end{align}
While for $m=m'$, we can make use of the fact that
\begin{align}
\mathbb{E}\left\{ n_{m,\mathrm{L}}^{2}(k)n_{m,\mathrm{L}}^{2}(k')\right\} & =  \mathbb{E}\left\{ n_{m,\mathrm{L}}^{2}(k)\right\}\mathbb{E}\left\{n_{m,\mathrm{L}}^{2}(k')\right\} \nonumber \\
&+2\left(\mathbb{E}\left\{ n_{m,\mathrm{L}}(k)n_{m,\mathrm{L}}(k')\right\} \right)^{2} \nonumber \\
 & =  \left(N_{0}W\beta\right)^{2}+2\left(N_{0}Wg(k-k')\right)^{2},
\end{align}
where $g(k)$ is a discrete low-pass filter with gain 1 in the band $[-\beta W/2, \beta W/2]$ and where we have utilized the fact that
\begin{eqnarray}
\mathbb{E}\left\{ n_{m,\mathrm{L}}(k)n_{m,\mathrm{L}}(k')\right\} =N_{0}Wg(k-k').
\end{eqnarray}
Hence,
\begin{align}
\psi(m,m) & =  \frac{1}{N_{0}^{2}\beta^{2}W^{2}N_{\mathrm{rep}}^{2}}\sum_{k,k'}\Bigl[\left(N_{0}W\beta\right)^{2} \nonumber \\& + 2\left(N_{0}Wg(k-k')\right)^{2}\Bigr]  \nonumber     \\
 & =  \frac{1}{N_{0}^{2}\beta^{2}W^{2}N_{\mathrm{rep}}^{2}} \Bigl[\left(N_{0}W\beta N_{\mathrm{rep}}\right)^{2}\nonumber \\
 & +2N_{0}^{2}W^{2}\sum_{k,k'}g^{2}(k-k')\Bigr] \nonumber \\
 & \approx  \frac{1}{N_{0}^{2}\beta^{2}W^{2}N_{\mathrm{rep}}^{2}} \Bigl[\left(N_{0}W\beta N_{\mathrm{rep}}\right)^{2} \nonumber \\
 & + 2N_{0}^{2}W^{2}N_{\mathrm{rep}}\sum_{k=-\infty}^{\infty}g^{2}(k)\Bigr] \nonumber \\
 & =  \frac{1}{N_{0}^{2}\beta^{2}W^{2}N_{\mathrm{rep}}^{2}}\left[\left(N_{0}W\beta N_{\mathrm{rep}}\right)^{2} + 2N_{0}^{2}W^{2}\beta N_{\mathrm{rep}}\right] \nonumber \\
 & =  1+2/(\beta N_{\mathrm{rep}}).
\end{align}
Substitution leads to
\begin{align}
\mathbb{E}\left\{ D^{2}(\tau^{*})\right\}  & =  \frac{1}{T_{\mathrm{p}}^{2}W^{2}}\left(\sum_{m}\psi(m,m)+\sum_{m,m\neq m'}\psi(m,m')\right)  \nonumber \\
 & =  \frac{1}{T_{\mathrm{p}}^{2}W^{2}}\Bigl(T_{\mathrm{p}}W\times\left(1+2/(\beta N_{\mathrm{rep}})\right)\nonumber \\
 &+\left(T_{\mathrm{p}}^{2}W^{2}-T_{\mathrm{p}}W\right)\times1\Bigr)  \nonumber \\
 & =  \frac{1}{T_{\mathrm{p}}^{2}W^{2}}\left(2T_{\mathrm{p}}W/(\beta N_{\mathrm{rep}})+T_{\mathrm{p}}^{2}W^{2}\right)  \nonumber \\
 & =  1+\frac{2}{T_{\mathrm{p}}\beta WN_{\mathrm{rep}}}.
\end{align}
So that we finally obtain
\begin{eqnarray}
\sigma_{\mathrm{f}}^{2}=\frac{2}{T_{\mathrm{p}}\beta WN_{\mathrm{rep}}}.
\end{eqnarray}

\subsection{Derivation of Statistics for Missed Detection Probability \label{subsec:Apendix2}}

We recall that in the presence of a person,
\begin{align}
D(\tau^{*}) & =\frac{1}{T_{\mathrm{p}}N_{0}\beta W^{2}N_{\mathrm{rep}}}\sum_{m=\left(\tau^{*}-T_{\mathrm{p}}/2\right)W}^{\left(\tau^{*}+T_{\mathrm{p}}/2\right)W}\Bigl(\left\Vert \mathbf{x}_{m}\right\Vert ^{2}+\left\Vert \mathbf{n}_{m,\mathrm{L}}\right\Vert ^{2}\nonumber\\
& +2\mathbf{x}_{m}^{T}\mathbf{n}_{m,\mathrm{L}}\Bigr) \nonumber \\
& =\frac{G(\tau_{\mathrm{p},1})E_s}{T_{\mathrm{p}}N_{0}\beta W^{2} N_{\mathrm{rep}}}\nonumber \\
& \times \sum_{m=\left(\tau^{*}-T_{\mathrm{p}}/2\right)W}^{\left(\tau^{*}+T_{\mathrm{p}}/2\right)W}V^{2}(m/W-\tau_{\mathrm{p},1})\left\Vert \mathbf{w}\right\Vert ^{2}\nonumber \\
& +\frac{1}{T_{\mathrm{p}}N_{0}\beta W^{2}N_{\mathrm{rep}}}\sum_{m=\left(\tau^{*}-T_{\mathrm{p}}/2\right)W}^{\left(\tau^{*}+T_{\mathrm{p}}/2\right)W}\left\Vert \mathbf{n}_{m,\mathrm{L}}\right\Vert ^{2} \nonumber \\
& +\frac{2}{T_{\mathrm{p}}N_{0}\beta W^{2}N_{\mathrm{rep}}}\sum_{m=\left(\tau^{*}-T_{\mathrm{p}}/2\right)W}^{\left(\tau^{*}+T_{\mathrm{p}}/2\right)W}\mathbf{x}_{m}^{T}\mathbf{n}_{m,\mathrm{L}}.
\end{align}
We easily find that
\begin{eqnarray}
\mathbb{E}\left\{ D(\tau^{*})\right\} =\frac{G(\tau_{\mathrm{p},1})E_{s}}{T_{\mathrm{p}}N_{0}\beta W}+1.
\end{eqnarray}
Recalling that odd-order moments of zero-mean Gaussian random variables are zero, we  determine $\mathbb{E}\left\{ D^{2}(\tau^{*})\right\} $ explicitly
as
\begin{align}
\mathbb{E}\left\{ D^{2}(\tau^{*})\right\}  & = \left(\frac{G(\tau_{\mathrm{p},1})E_{s}}{T_{\mathrm{p}}N_{0}\beta W}\right)^{2}\nonumber \\
& +\mathbb{E}\left\{\left(\frac{1}{T_{\mathrm{p}}N_{0}\beta W^2N_{\mathrm{rep}}}\sum_{m}\sum_{k}n_{m,\mathrm{L}}^{2}(k)\right)^{2}\right\} \label{eq:bigexpansion}\nonumber\\
 & +  \mathbb{E}\left\{\left(\frac{2}{T_{\mathrm{p}}N_{0}\beta W^2 N_{\mathrm{rep}}}\sum_{m}\sum_{k}n_{m,\mathrm{L}}(k)x_{m}(k)\right)^{2}\right\} \nonumber \\
 & +  \frac{2G(\tau_{\mathrm{p},1})E_{s}}{T_{\mathrm{p}}^{2}N_{0}^{2}\beta^{2}W^{3}N_{\mathrm{rep}}}\sum_{m}\sum_{k}\mathbb{E}\left\{ n_{m,\mathrm{L}}^{2}(k)\right\} .
\end{align}
Here the second term simplifies to $1+2/(T_{\mathrm{p}}\beta WN_{\mathrm{rep}})$, the fourth term to $2G(\tau_{\mathrm{p},1})E_s/(T_{\mathrm{p}}N_0\beta W)$. The third term is equal to
\begin{align}
\mathrm{Term_{3}} & = \left(\frac{2}{T_{\mathrm{p}}N_{0}\beta W^2N_{\mathrm{rep}}}\right)^2\sum_{m,m'}\sum_{k,k'}\mathbb{E}\left\{ n_{m,\mathrm{L}}(k)n_{m',\mathrm{L}}(k')\right\}\nonumber \\
&\times \mathbb{E}\left\{x_{m}(k)x_{m'}(k')\right\} \nonumber\\
 & = \left(\frac{2}{T_{\mathrm{p}}N_{0}\beta W^2 N_{\mathrm{rep}}}\right)^2\left(\sum_{m}A(m)+\sum_{m, m'\neq m}B(m,m')\right),
\end{align}
in which
\begin{align}
A(m) & =  \sum_{k}\sum_{k'}\mathbb{E}\left\{n_{m,\mathrm{L}}(k)n_{m,\mathrm{L}}(k')\right\}\mathbb{E}\left\{ x_{m}(k)x_{m}(k')\right\}  \nonumber\\
 & =  N_{0}W\sum_{k}\sum_{k'}g(k-k')\mathbb{E}\left\{x_{m}(k)x_{m}(k')\right\} \nonumber \\
 & =  N_{0}W\sum_{k}\mathbb{E}\left\{x_{m}^{2}(k)\right\},
\end{align}
and $B(m,m')=0$. Hence,
\begin{align}
\mathrm{Term_{3}} & =  \frac{4N_0W}{T_{\mathrm{p}}^2N_{0}^2\beta^2 W^4 N_{\mathrm{rep}}^2} \sum_{m}\sum_{k}\mathbb{E}\left\{x_{m}(k)^2\right\} \nonumber \\
 & = \frac{4N_0WG(\tau_{\mathrm{p},1})E_s}{T_{\mathrm{p}}^2N_{0}^2\beta^2 W^4 N_{\mathrm{rep}}^2}\sum_{m}\sum_{k}V^2(m/W-\tau_{\mathrm{p},1})\mathbb{E}\left\{w^{2}(k)\right\} \nonumber \\
 & = \frac{4G(\tau_{\mathrm{p},1})E_s}{T_{\mathrm{p}}^2N_0\beta^2 W^2 N_{\mathrm{rep}}}.
\end{align}
Putting everything together, we find
\begin{align}
\mathbb{E}\left\{ D^{2}(\tau^{*})\right\}& =\left(\frac{G(\tau_{\mathrm{p},1})E_{s}}{T_{\mathrm{p}}N_{0}\beta W}\right)^{2}+1+\frac{2}{T_{\mathrm{p}}\beta WN_{\mathrm{rep}}}\nonumber \\
&+\frac{G(\tau_{\mathrm{p},1})E_{s}}{T_{\mathrm{p}}\beta W N_{0}}\left(2+\frac{4}{T_{\mathrm{p}}\beta WN_{\mathrm{rep}}}\right).
\end{align}
Subtracting
\begin{eqnarray}
\mathbb{E}^2\left\{ D(\tau^{*})\right\}  = 1+\frac{2G(\tau_{\mathrm{p},1})E_{s}}{T_{\mathrm{p}}N_{0}\beta W}+\left(\frac{G(\tau_{\mathrm{p},1})E_{s}}{T_{\mathrm{p}}N_{0}\beta W}\right)^{2},
\end{eqnarray}
gives us
\begin{eqnarray}
\sigma_{\mathrm{d}}^{2} = \frac{2}{T_{\mathrm{p}}\beta WN_{\mathrm{rep}}}\left(1+ \frac{4G(\tau_{\mathrm{p},1})E_s}{T_{\mathrm{p}}N_0\beta W}\right).
\end{eqnarray}

\end{document}